# The Need for a New Generation of Space-based Visible and Near-IR Emission Line Observations of the Corona

A White Paper for the 2024 Solar and Space Physics (Heliophysics) Decadal Survey

Benjamin Boe[1], Shadia Habbal[1], and Adalbert Ding[2,3]

1. Institute for Astronomy, University of Hawaii
2. Institute of Optics and Atomic Physics, Technische Universitaet Berlin, Germany
3. Institut fur Technische Physik, Berlin, Germany

## Synopsis

Visible and near-infrared (V+NIR) emission lines were the first to be discovered in the corona, during total solar eclipses, and they continue to offer unique opportunities to study the physical properties of the corona. The most commonly observed coronal emission lines today are in the extreme ultraviolet, which are dominated by collisionally excited emission. V+NIR lines on the other hand, are radiatively excited out to high helioprojective distances. Indeed, recent eclipse observations have demonstrated the diagnostic potential of V+NIR lines, which are still observable out to at least 3.4 $R_\odot$. V+NIR lines can be used to infer key plasma parameters such as: the electron and ion temperatures, magnetic field strength and morphologies, the ionic freeze-in distances, Doppler motions of coronal plasmas, and the dynamics of coronal mass ejections through time variations of these parameters. Current and planned space-based coronagraphs, such as Solar Orbiter and Proba 3, will have some filters for V+NIR lines, but will only have an exceptionally small selection. They will thus be limited in their ability to infer electron or ion temperatures, as well as other crucial physical properties of the corona. The ground-based DKIST and UCoMP will soon offer V+NIR line observations, but they will be limited to a maximum helioprojective distance of 1.5 and 2 $R_\odot$ respectively, in large part due to the brightness of the Earth's atmosphere. To better explore the middle (~1.5–6 $R_\odot$) and outer corona (> 6 $R_\odot$), and to understand the formation of the solar wind and space weather events, it is essential in this next decade that we **deploy additional space-based assets to measure a wide selection of V+NIR emission lines at helioprojective distances beyond 1.5 $R_\odot$, with both spectrometers and narrowband imagers.** Occulting of the solar disk could be achieved not only by a conventional coronagraph, but also by novel methods such as an external occulter, by lunar occultations in situ in orbit around the Moon, or by lunar based observations of lunar eclipses (i.e., total solar eclipse on the Moon).

# 1. Historical Context

The first discovery of a coronal emission line occurred with a slit-less spectroscopic experiment during the 1869 Total Solar Eclipse (TSE; Young 1872). The so-called "green" coronal emission line was later realized (Grotrian 1939, Eldén 1943) to be emission from the highly ionized $Fe^{13+}$ ion (or the Fe XIV line in astronomical notation). With the invention of the coronagraph by Lyot in the 1930s, emission lines could then be observed outside of TSEs (e.g., Lyot 1939).

Coronagraph observations of the Fe XIV (530.3 nm) line followed over several decades in the 20th century (e.g., Rybansky et al. 1994, Altrock 2011), albeit in the low corona (below 1.2 $R_\odot$). There were also limited observations of the Fe X (637.4 nm) and Fe XIV lines with the space-based coronagraph LASCO-C1 (e.g., Mierla et al. 2008). Unfortunately, LASCO-C1 failed shortly after launch and did not have a long enough period of data collection to demonstrate the full potential of such observations. Even so, the data were sufficient to measure the line emission and line width of Fe X and Fe XIV (see Figure 1). LASCO-C1 was in fact the last space-based coronagraph observation to observe V+NIR emission lines (until Solar Orbiter and Proba 3, see Section 3).

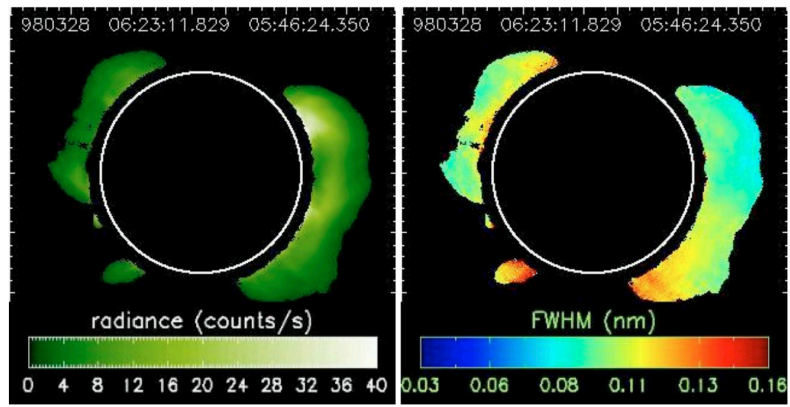

Figure 1: Fe XIV radiance (left), and line width (right) from the LASCO-C1 coronagraph (from Mierla et al. 2008)

At present, most of the coronal research community has been focusing on space-based extreme ultraviolet (EUV) observations. V+NIR wavelengths are now almost exclusively observed via broadband continuum white-light (including polarization) filters, where emission originates from electrons and dust (e.g., Lamy et al. 2020). **The white-light observations are useful for measuring the electron density in the corona, but offer no method for studying other properties of the corona, such as electron or ion temperatures, magnetic field strength, ionic freeze-in distances or Doppler motions** (V+NIR lines can, see Section 2).

The EUV observations have returned a vast volume of valuable data in the lower corona, but they are inherently limited due to their predominantly collisionally excited nature. That is, EUV lines are excited mostly by collisions in the corona, since the solar disk does not emit EUV photons in any meaningful quantity. Hence the brightness of EUV lines is proportional to the density squared and so the brightness drops dramatically as the expanding corona decreases by orders of magnitude in density. It becomes very tenuous beyond about 1.5 $R_\odot$, and is effectively not observable beyond about 2.5 to 3 $R_\odot$ (see Figure 2, Seaton et al. 2021).

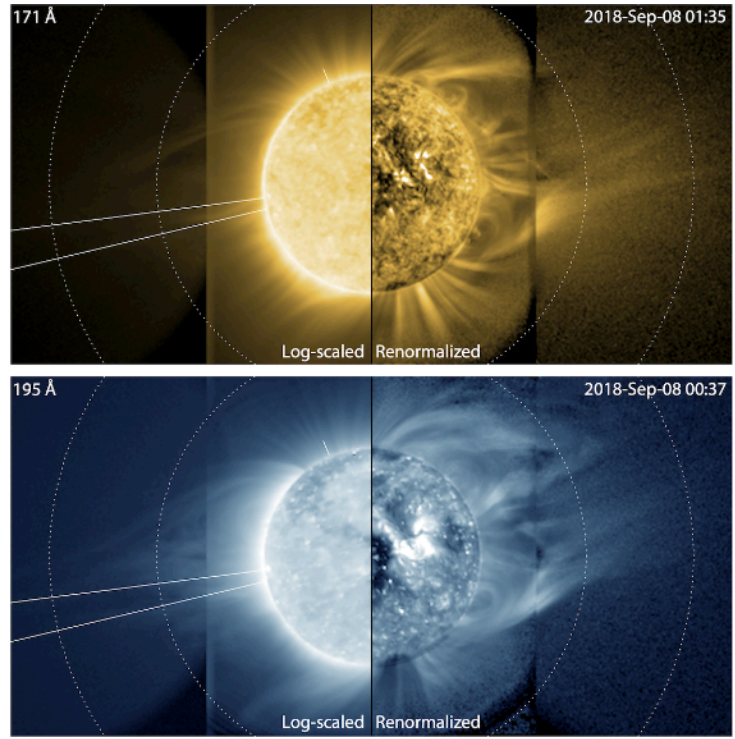

Figure 2: EUV 17.1 nm (top), and 19.1 nm (bottom) emission from GOES-SUVI. The left portion of each panel is in a traditional log-scaling, whereas the right has been processed to show detail of coronal structures (from Seaton et al. 2021).

There is still some radiative excitation of EUV lines farther out in the corona, caused by emission from the same line collisionally excited lower down. Such observations of EUV lines farther out are then subject to a variable incident flux that makes any analysis significantly more challenging. Finally, the close spacing of lines in the EUV means that bandpasses, such as those on SDO/AIA, are cross-contaminated by multiple lines leading to exceptionally complex temperature response functions for each observation (see Boerner et al. 2012).

V+NIR emission lines, on the other hand, do not have any of the same limitations that the EUV lines do. The solar disk supplies ample photons for radiative excitation, which is proportional to the density (not squared). V+NIR emission lines can thus be observed at much higher helioprojecitve distances than can EUV lines, and they are spaced out in wavelength such that there is no cross contamination between spectral line observations.

Clearly, the EUV emission observations are essential for probing the low corona (below about 1.5 to 2 $R_\odot$), but the middle corona (broadly defined as 1.5 to 6 $R_\odot$) is actually much better probed by V+NIR emission lines. However, the limitations of sky-brightness for ground-based observations requires that we deploy a new generation of space-based observations of the corona targeting a wide range of V+NIR lines in this next decade (see Section 3).

## 2. The Diagnostic Potential of Visible and NIR Emission Lines

Despite the lack of space-based V+NIR observations since LASCO-C1 failed (see Section 1), there has still been a thriving application of these emission lines with TSEs and ground-based coronagraphs that have demonstrated the potential of such observations. For example, recent ground-based coronagraph observations of Fe XIII (1074.7 nm) have been used to infer the coronal magnetic field direction and strength (e.g., Lin et al. 2004, Gibson et al. 2017), and Alfvén waves in the corona (Tomczyk et al. 2007).

Recent TSE data have further demonstrated the potential of V+NIR spectral lines, through observations of the Fe X, Fe XI (789.2 nm), and Fe XIV. These lines cover the temperature range expected in the corona quite well, so they can be used to infer the spatially resolved coronal electron temperature (Te; Boe et al. 2022, Figure 3). The Fe XI (Fe X) line is also observable to at least 3.4 (3) $R_\odot$ during an eclipse, which only lasts a couple of minutes. It is therefore likely that space-based observations could observe these lines at even higher elongations (see Section 4).

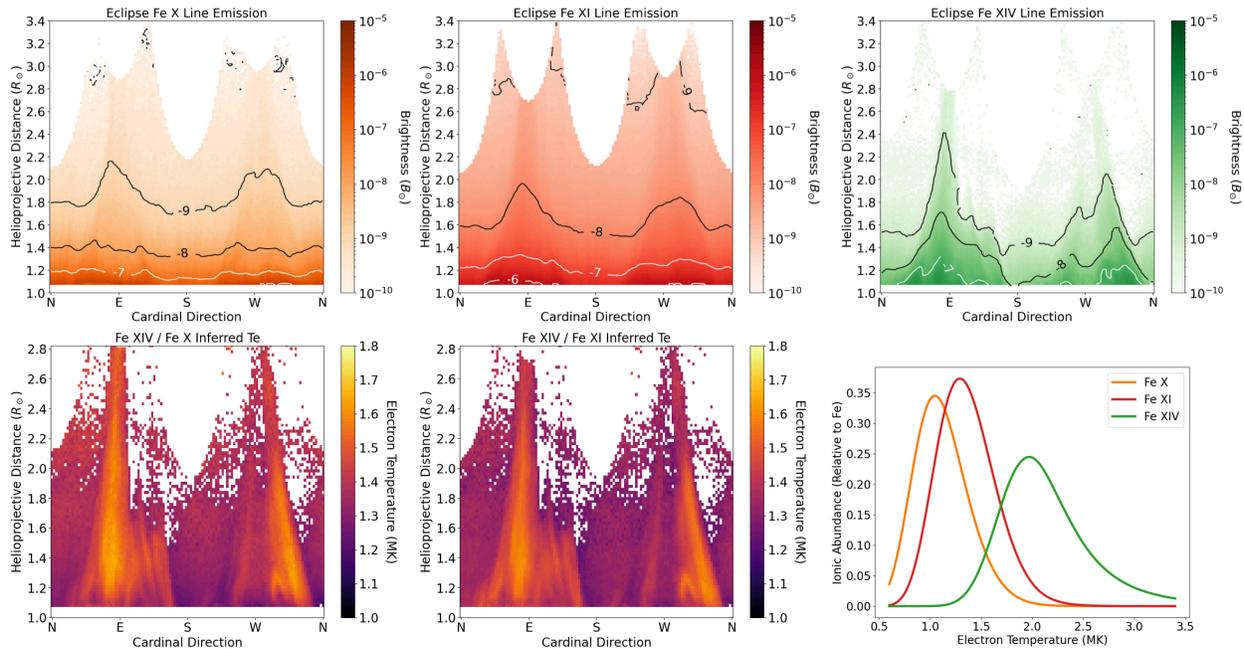

Figure 3. Top row: Absolute brightness of Fe X (left), Fe XI (middle) and Fe XIV (right). Bottom row: Inferred line ratio Te from Fe XIV/Fe X (left) and Fe XIV/Fe XI (middle) based on the ionization temperature curves (right) from CHIANTI (from Boe et al. 2022).

Imaging observations of these lines are even more powerful when there is a time baseline between observations to measure changes, as would be the case with space-based observations. For example, during the 2017 TSE across the United States, Boe et al. (2020) was able to use identical Fe XI and Fe XIV observations between Oregon and Nebraska to measure Te changes in the corona (line of sight average) over 28 minutes due to a propagating halo-CME in the corona (Figure 4).

An important feature of V+IR line observations is that the continuum emission is as strong as the lines, thus requiring additional continuum observations at nearby wavelengths (within ~3-5 nm). However, the continuum data are independently useful, as the relative color of the dust (F corona) and electron (K corona) scattering can be separated using a color-based method (Boe et al. 2021). The ratio between the continuum and line emission can also be used to infer the ionic freeze-in distances (Habbal et al. 2007, Boe et al. 2018, Figure 5), which offers an empirical method for linking coronal structures to their fingerprint in the solar wind (see Habbal et al. 2010, 2021).

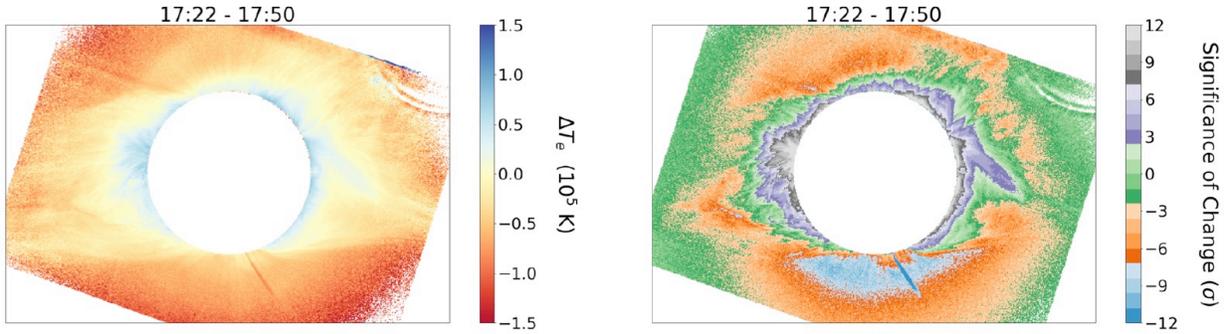

Figure 4. Change in the inferred Te over 28 minutes between Mitchell, OR and Alliance, NE during the 2017 TSE. The left panel shows the total change, the right panel shows the statistical significance of the change (from Boe et al. 2020).

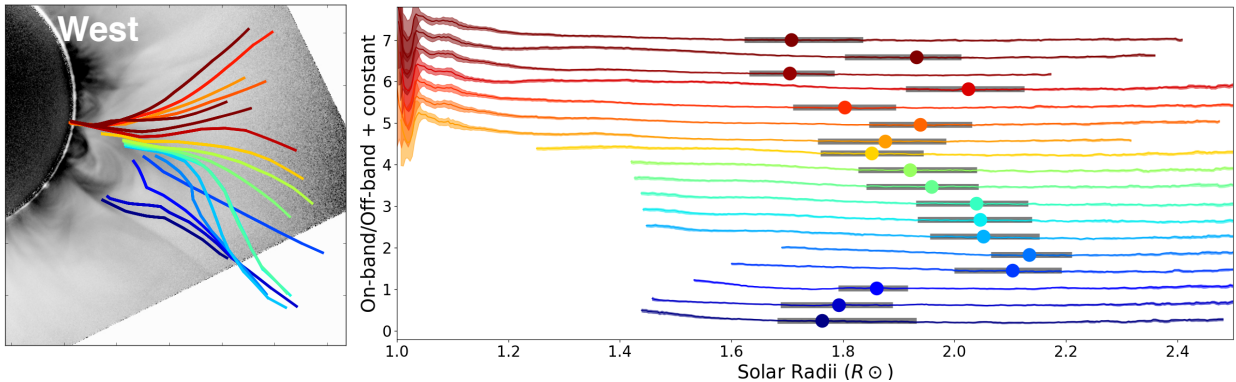

Figure 5. Demonstration of the Freeze-in distance measurement for Fe XIV (from Boe et al. 2018).

In addition to narrowband imaging observations, slit-spectroscopic observations during TSEs have also demonstrated the potential of spectroscopic observations of V+NIR lines. Observations from the 2015 TSE led to the detection of large Doppler shifted motions of Fe XIV and chromospheric lines (from prominence material) which had been launched into the corona by a coronal mass ejection (CME; Ding & Habbal 2017, Figure 6). Spectroscopic eclipse observations can also identify various heavy element species (e.g., Samra et al. 2018, Koutchmy et al. 2019), which could be used to measure elemental abundances and ergo elemental variations caused by the first ionization potential (FIP) effect (see Schmelz et al. 2012) — especially with an uninterrupted temporal coverage.

## 3. Near Term Outlook

It is clear from Section 2 that spectroscopic and narrowband imaging observations in the V+NIR are powerful tools to probe dynamics in the corona and the formation of the solar wind. Observations of V+NIR lines from the ground will soon be expanded with UCoMP, which will be able to produce many similar studies to what has been done during TSEs (see Figure 7), and will add the polarization of the lines for magnetic field inferences (see Section 2). However, UCoMP will be limited to a maximum of 2 $R_\odot$, and it is unlikely that it will be able to regularly achieve robust signals above 1.5 $R_\odot$.

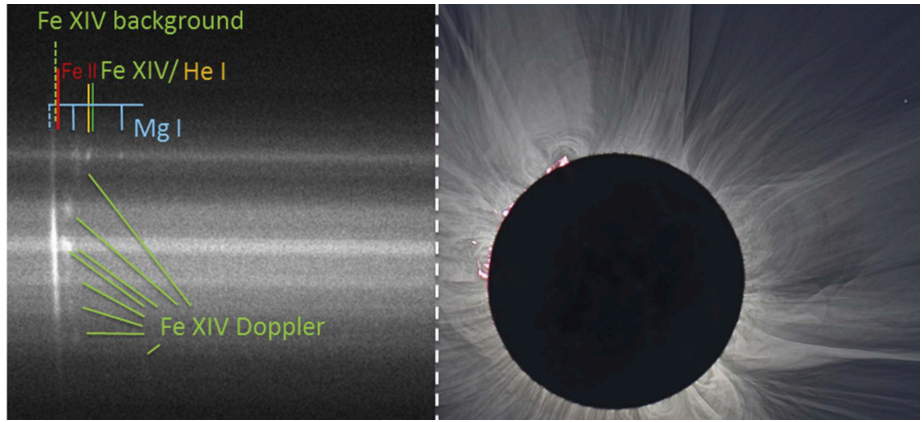

Figure 6. Spectroscopic observations from the 2015 TSE. The left panel shows a single spectrum with the slit aligned on the dashed line in the white-light image to the right. The spectrum shows Fe XIV emission and Doppler shifted emission from a CME, along with low ionization state lines from erupting prominence material (from Ding & Habbal 2017).

DKIST will also be able to study V+NIR emission lines in the corona using DL-NIRSP and Cryo-NIRSP, but it will be limited in its pointing to 1.5 $R_\odot$ at most. Consequently, DKIST data will be useful for studying small-scale dynamics in the low corona, but it is incapable of measuring emission from the middle corona.

The combination of DKIST, UCoMP, and EUV observations will cover the low corona with a wide range of diagnostic potential over the coming years. Nevertheless, the middle corona will still be largely unexplored by these data. Space-based coronagraphs such as Solar Orbiter and Proba 3 will begin to perform some narrowband observations in the middle corona, but together they will only get Lyman Alpha (H I), He I and Fe XIV. These observations will be valuable for comparison with other observations from the ground, but are insufficient for measuring electron or ion temperatures. They will correspondingly be quite limited in their diagnostic potential.

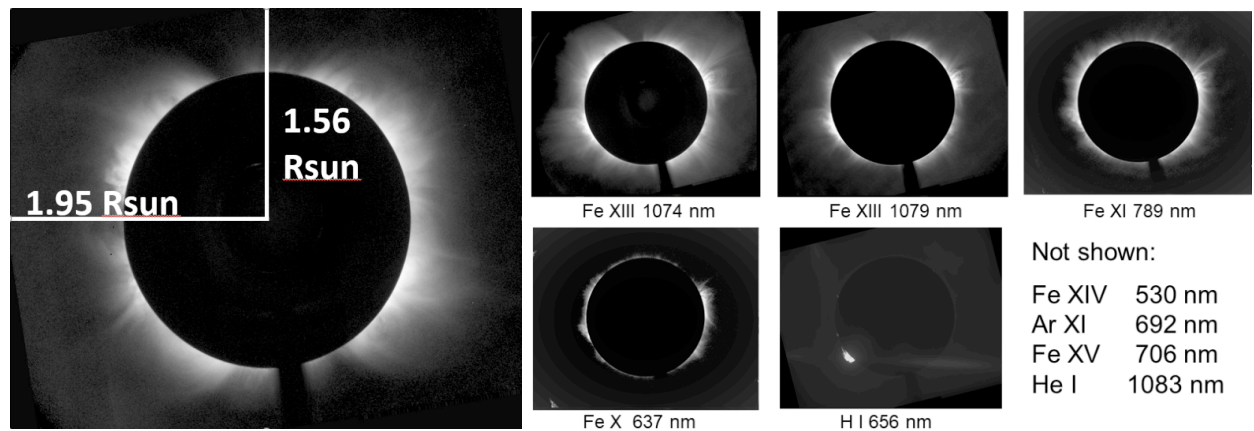

Figure 7. Early Fe XIII brightness observation from UCoMP (left) showing the field of view of the instrument. The right panels show a selection of other lines observed by UCoMP (from UCoMP website).

The best opportunities at present to study a vast range of properties in the middle corona remains to be TSEs, as it has for over a century. The TSE data will continue to provide insightful data on the corona no doubt, but these studies are inherently limited by the rarity of TSE occurrence on the Earth. New space-based observatories, focusing on V+NIR emission lines from 1.5 to > 6 $R_\odot$ are therefore essential in the next decade to advance our understanding of the middle corona and the formation of the solar wind.

## 4. Proposed Future Observations

Observations of V+NIR lines have a long history going back to the 19th century (see Section 1), and recent observations of this type have returned a plethora of information on the corona and solar wind (see Section 2), but there are no existing nor planned observations — outside of eclipses — which will be able to achieve sufficient diagnostic capabilities beyond about 1.5 to 2 $R_\odot$ (see Section 3).

Since V+NIR lines have now been observed as far as 3.4 $R_\odot$ during a TSE (see Figure 3), it is clear that the **line emission could be measured at much higher helioprojective distances than previously imagined**, if one had the correct instrumentation to do so. Thus, we need to deploy additional space-based coronagraph facilities that can measure a wide range of emission lines throughout the middle corona. The last attempted observations of V+NIR emission lines from space was LASCO-C1, which launched in 1995. With the advancement of detectors and spacecraft technology over that interim period, it is reasonable to expect that one could achieve substantially better performance with a new instrument of the same type. Further, the finding of emission at 3.4 $R_\odot$ means that even LASCO-C2 (i.e., 2 to 6 $R_\odot$) may very well have been able to detect line emission, if it had been equipped with the requisite filters. A new version of a LASCO-C2 type instrument equipped both with white light polarization and narrowband filters would be of extreme benefit for future studies of the middle corona.

Specifically the V+NIR Fe lines of Fe X, Fe XI, Fe XIII, Fe XIV, Fe XV, and the Ar lines of Ar X, Ar XI, as well as lines from many other species (especially of Ni, Si and Ca) could all be measured with similar coronagraph design to C2 and a filter wheel with narrow bandpasses. The Fe XI emission line in particular has never been observed from space, yet it is found to be the brightest emission line during TSEs (Boe et al. 2022) and the most abundant ion in the solar wind (Habbal et al. 2010, 2021). Having a large set of emission lines would also enable inferences of the resolved electron temperature distribution along each line-of-sight in the corona (via a DEM type method, see Del Zanna & Mason 2018), measure the time variable and temperature dependence of the freeze-in distance (see Boe et al. 2018), study the thermodynamics of CME evolution in the corona (see Boe et al. 2020), and measure the FIP effect using lines from elements with different first ionization potentials.

The requisite continuum bandpass images for each line would additionally offer a means to independently extract the K and F corona without polarization (see Boe et al. 2021). Further, these continuum images would help to reduce any stray light contamination in the line emission data, since the stray light would be effectively the same over a small wavelength shift between the line and continuum filters.

That same set of emission lines, and even more from various heavy elements, should simultaneously be explored with V+NIR spectrometers to study the line-widths, Doppler motions, and Alfvénic wave disturbances throughout the middle corona. UVCS had shown the potential of such observations in the UV (e.g., Esser 1999), but V+NIR emission lines would be even more powerful diagnostic tools moving forward to link the corona and solar wind.

A space-based coronagraph of the type outlined here could likely be done on a SMEX or MIDEX level mission proposal, as one could use largely heritage design and instrumentation, with the only new addition being a filter wheel capable of observing a wide range of V+NIR lines. A Medium-class or Large-scale mission would be even more valuable, as one could add additional instrumentation for in situ space plasma instruments to measure ionic composition (e.g., SWICS) and various plasma parameters and could even have multiple identical spacecraft that could be sent in different heliocentric orbits, enabling multiple perspectives (similar to the STEREO mission). There could also be deployment of an external occulter spacecraft, similar to what Proba 3 will be testing in the near future. If that technology is demonstrated to work effectively, then a subsequent mission could use the same method with a wider range of V+NIR line capabilities.

One promising future development in the next decade that could assist in coronal observations, will be regular manned **space missions to the lunar surface and the planned Lunar Gateway orbital station**. Some portion of the moon experiences a TSE (i.e., lunar eclipse where the Earth is the occulter) at least once every 6 months (sometimes twice) compared to an average of ~18 months between TSEs on the Earth. With missions to the Moon, we as a community should plan to deploy instrumentation to observe the corona during the frequent eclipses. As the Earth moves from one side of the corona to the other, the observations could focus on the opposite side of the corona for half of the eclipse duration (which is hours on the Moon, compared to minutes on the Earth). Rather small payloads, similar to the ones used during eclipses on the Earth, could be deployed to make regular observations during lunar eclipses. A mission of this sort could perhaps be done with as little as a Missions of Opportunity scale by hitching a ride on a larger-scale lunar mission.

Another similar idea was proposed by Habbal et al. (2013), where they suggested that a lunar orbiting mission could use the moon itself as an occulter to observe half of the corona during sunrise/sunset of the orbiting craft. This scenario combined with the lunar eclipses could offer frequent deep exposure data on these radiatively excited V+NIR emission lines without any coronagraph at all, and offer exceptionally valuable data for understanding the formation and evolution of the corona and solar wind. Again, this mission could hitch a ride on a larger mission to deployment in a lunar orbit which would reduce the price to perhaps a SMEX level mission.

Nevertheless, TSE observations should still remain a fixture of coronal observations (see White Paper by Habbal et al. 2022), especially as a testing ground for novel instrumentation. The nature of TSE observing allows one to build new and improve older instrumentation methods, including searching for emission lines that have not previously been studied in detail. Such TSE testing can be used for spacecraft concepts on the ground before they are launched.

In this next decade, we should strive to expand the diversity of coronal observations including all wavelengths from the X-Ray to the Radio, but specifically space-based observations of the V+NIR. The relative dearth of current and planned V+NIR observations in the middle corona specifically is a major impediment to the development of our understanding of the physical processes throughout the corona and thus the sources of the solar wind and CMEs.